\documentclass[showpacs,twocolumn,aps,preprintnumbers,letterpaper]{revtex4}
\usepackage{amsmath,amssymb}
\usepackage{epsfig}
\usepackage{graphicx}
\usepackage{amsmath}
\usepackage{slashed}
\usepackage{amsfonts}
\usepackage{epstopdf}
\usepackage{color}

\newcommand{\be}{\begin{equation}}
\newcommand{\ee}{\end{equation}}
\newcommand{\bear}{\begin{eqnarray}}
\newcommand{\eear}{\end{eqnarray}}
\newcommand{\ba}{\begin{array}}
\newcommand{\ea}{\end{array}}

\def\be{\begin{eqnarray}}
\def\ee{\end{eqnarray}}

\def\roughly#1{\mathrel{\raise.3ex\hbox{$#1$\kern-.75em%
\lower1ex\hbox{$\sim$}}}}
\def\lsim{\roughly<}

  \long\def\comment#1{ }

  \newcommand{\rmd}{{\rm d}}   

  \newcommand{\nn}{\nonumber\\}

  \newcommand{\calbfA}{{\bm{\mathcal{A}}}}

  \newcommand{\beq}{\begin{eqnarray}}
  \newcommand{\eeq}{\end{eqnarray}}
  
 \def\simge{\mathrel{%
   \rlap{\raise 0.511ex \hbox{$>$}}{\lower 0.511ex \hbox{$\sim$}}}}
\def\simle{\mathrel{
   \rlap{\raise 0.511ex \hbox{$<$}}{\lower 0.511ex \hbox{$\sim$}}}}

\begin{document}

\title{Deep Inelastic Scattering on an Extremal RN-AdS Black Hole}

\author{Kiminad A. Mamo and Ismail Zahed}
\email{kiminad.mamo@stonybrook.edu}
\email{ismail.zahed@stonybrook.edu}
\affiliation{Department of Physics and Astronomy, Stony Brook University, Stony Brook, New York 11794-3800, USA}



\date{\today}
\begin{abstract}
We consider deep inelastic scattering (DIS) 
on a large nucleus described as an extremal  RN-AdS black hole using the holographic principle.
Using the R-current correlators we determine the structure functions  as a function Bjorken-x, and
map it on a finite but large nucleus with fixed atomic number.  The R-ratio of the nuclear structure functions 
exhibit strong shadowing at low-x.
\end{abstract}


\maketitle

\setcounter{footnote}{0}


\section{Introduction}

DIS scattering on nuclei have shown that the nuclear structure functions 
deviate substantially from that of the nucleon, especially at low Bjorken-x~\cite{E665,NMC,SLAC,BCDMS}. 
The depletion at low-x is usually refered to as shadowing. It is currently  understood as the coherent scattering
on two or more nucleons in the nucleus,  as opposed to incoherent scattering on individual nucleons. In this sense,
shadowing maybe reminiscent of diffractive scattering in high energy hadron-on-hadron scattering.

At extremely low-x, the measured nucleon structure function shows a rapid growth of 
partons, primarily gluons~\cite{H1,ZEUSS}. Phenomenological arguments suggest that
the growth saturates~\cite{GW}, a point supported by perturbative QCD arguments~\cite{BK}. 
A central question is then: how is the growth of low-x 
partons in a nucleon, realized in a nucleus? Is shadowing further enhanced at low-x?
Some of these important questions will be addressed and hopefully answered in the future electron-Ion-Collider (eIC).

DIS in holography at moderate-x is different from weak coupling as it involves hadronic  and
not partonic constituents~\cite{POL}. The large gauge coupling causes the charges to rapidly deplete
their energy and momentum, making them invisible to hard probes. 
However, because the holographic limit enjoys approximate conformal
symmetry,  the structure functions and form factors exhibit various scaling laws including the 
parton-counting rules~\cite{BF}. In contrast, DIS scattering at low-x on a non-extremal
thermal black-hole was argued  to be partonic  and fully saturated~\cite{HATTA}.

In this paper we consider DIS  scattering on a large but finite nucleus in holography 
using  an extremal RN-AdS black-hole~\cite{adsBH} in the conformal limit. We use the mapping
of the RN-AdS charge on the atomic number  to construct the nuclear R-ratio.  This point of view takes to
the extreme the concept of coherent DIS scattering on a dense nucleus, and therefore should
be of relevance in the shadowing or low-x region. The intermediate-x and large-x regions are
subleading in the holographic limit as we explain below and detail elsewhere. Our
arguments will be similar to those presented for the thermal non-extremal black-hole in~\cite{HATTA},
with the key difference being the large but finite charge as the atomic number.

The organization of the paper is as follows: in section II we briefly review the setting for the 
RN-AdS black hole, and specialize to the extremal case with zero temperature.  In section III
we detail DIS scattering on the extremal black hole in leading order in the holographic limit,
and give the pertinent on-shell action for the probe gauge field. In section IV we explicit
the holographic structure functions on nucleus as an extremal RN-AdS black hole in the conformal limit.  The
R-ratio is constructed and shown to display shadowing at low-x. Our conclusions are in section V.

\section{RN-AdS Black-Hole}

 Studies of fermionic systems in the context of gravity dual theories 
have been carried by many~\cite{FIRST,SIN,MANY}.
For dense nuclei we may simplify the nucleus by treating it as cold black hole in the conformal limit, using 
the holographic dual construction. In this section, we briefly review the essentials of an RN-AdS black hole,
and then specialize to the extremal case.

\subsection{General RN-AdS}

The effective action describing bulk RN-AdS gravity sourcing a U(1) gauge field reads

  \be
  S=\frac 1{2\kappa^2}\int d^5x\sqrt{-g}\,(\mathbb R-2\Lambda) -\frac 1{4e^2}\int d^5x\sqrt{-g}F^2
  \label{RNADS}\nonumber\\
  \ee
  with $\mathbb R$ the Ricci scalar, $\kappa^2=8\pi G_5$ and $\Lambda=-6/R^2$ are the gravitational and cosmological
  constant.   The ensuing gravitational equation is coupled to Maxwell equation 
  
  \be
 && R_{mn}-\frac 12 g_{mn}(\mathbb R -\Lambda)=\kappa^2T_{mn}\nonumber\\
 && T_{mn}=g^{pq}F_{mp}F_{qn}-\frac 14 g_{mn}F^{pq}F_{pq}\nonumber\\
 &&\frac 1{\sqrt{-g}}\partial^m\left(\sqrt{-g}\,F_{mn}\right)=0
 \label{GRA}
 \ee
Since $T^m_m=0$, the space is photon filled but with the  curvature of the AdS space. The RN-AdS 
black-hole solution to (\ref{GRA}) is charged in bulk with a U(1) scalar potential

  \be
  A_t=\mu-\frac {\mathbb Q}{r^2}
  \label{GA}
  \ee
  and a line element 
  
   \be
  ds^2=\frac {r^2}{R^2}\left(-f\, dt^2+d\vec{x}^2\right)+\frac {R^2}{r^2 f}dr^2
  \label{MADS}
  \ee
with 

\be
\label{FADS}
f=1-\frac {mR^2}{r^4}+\frac{q^2R^4}{r^6}
\ee
provided that  the electric charge $\mathbb Q$ and the geometrical charge $q$ satisfy

  \be
  \frac{q^2R^2}{\mathbb Q^2}=\frac 43 \times\frac{2\kappa^2}{4e^2}=\frac{R^2}{6\alpha}\,,
  \label{CHARGE}
  \ee
The last equality follows from the brane-filling set up, where the parameters can be identified as

 \begin{eqnarray}
   2\kappa^2=\frac{8\pi^2 R^3}{N_c^2}\qquad
   4e^2=\alpha\frac{64\pi^2 R}{N_c^2}
   \label{EOS}
   \end{eqnarray}
with $\alpha=1$ for a U(1) R-charge, and $\alpha=\frac{1}{4}\frac{N_c}{N_f}$ for a D3-D7  U(1) vector charge.


  \subsection{Extremal RN-AdS}

The RN-AdS black hole carries two horizons $f(r_\pm)=0$ with $r_+>r_-$, which are best seen by re-writing 
the warping factor (\ref{FADS}) as

 \be
 \label{FR}
 f(r)=\left(1-\frac{r_{+}^2}{r^2}\right)\left(1-\frac{r_{-}^2}{r^2}\right)\left(1+\frac{r_{+}^2}{r^2}+\frac{r_{-}^2}{r^2}\right)
 \ee 
with $mR^2=r_{+}^4+r_{-}^4+r_{+}^2r_{-}^2$ and $q^2R^4=r_{+}^2r_{-}^2(r_{+}^2+r_{-}^2)$,
provided that the mass $m$ and the geometrical charge $q$ satisfy  $q^4R^4\leq 4m^3R^2/27$, with  $R\sqrt{m/3}\leq r_+^2\leq R\sqrt{m}$.  The temperature of
the RN-ADS black hole is fixed by the standard requirement of no conical singularity in the vicinity of the outer horizon
$r_+$ 

  \be
  T=\frac{r_+^2f'(r_+)}{4\pi R^2}=\frac{r_+}{\pi R^2}\Big(1-\frac{\mu^2\pi^2 R^4\gamma^2}{r_+^2}\Big)
  \label{TEMP}
  \ee
with  $\gamma^2=\frac{1}{12\pi^2\alpha}$. Its  chemical potential $\mu$ is fixed by the zero potential condition 
on the outer horizon $A_t(r_+)=0$ or $\mu=\mathbb Q/r_+^2$. With these identifications,  the standard thermodynamics 
typical of black-holes follows.

The regulated Gibbs energy $\Omega=T\Delta S$ follows from (\ref{RNADS}) by inserting
  the RN-AdS charged black hole (\ref{MADS}-\ref{GA}) and subtracting the {\it empty} thermal
  AdS contribution~\cite{SIN}. The result is

   \be
   \Omega=-\frac{V_3}{2\kappa^2R^3}\left(\frac{r_+^4}{R^2}+\frac{q^2R^2}{r_+^2}\right)
   \label{OMEGA}
   \ee
   by trading $m=r_+^4/R^2+q^2R^2/r_+^2$. The entropy  $s$, energy $\epsilon$, pressure $p$ 
   densities and density $n$ follow from (\ref{OMEGA}) through the usual grand-canonical rule~\cite{adsBH,SIN}
   
   \begin{eqnarray}
  && s=\frac{2\pi r_+^3}{\kappa^2R^3}\nonumber\\
  && \epsilon=\frac{3m}{2\kappa^2R^3}=3p\nonumber\\
   &&n=\frac{2\mathbb Q}{e^2R^3}
   \label{EOS1}
   \end{eqnarray}
We will mostly consider the extremal RN-AdS black hole for which $T=0$ with $r_+=r_-=\pi R^2\gamma\mu$. Specifically, 
the bulk thermodynamical quantities in (\ref{EOS1}) simplify

\be
\label{EOS2}
 && s=\frac{2\pi}{\sqrt{3}}\sqrt{\alpha}n\nonumber\\
  && \epsilon=\frac 34 n\mu=3p\nonumber\\
   &&n=\frac {N_c^2}{96\pi^2\alpha^2}\mu^3
\ee

We  identify 
the extremal RN-AdS black hole with a very large but finite nucleus of volume 
$V_A=\frac 43 \pi R_A^3$ with a radius $R_A=R_1A^{\frac 13}$,
 a number density $A/V_A=n$, energy density $E_{A}/V_{A}=\epsilon$, and an energy per particle $E_A/A=\frac{3}{4}\mu$
(conformal). For comparison, nuclear matter with small scattering lengths carries  $E_A/A\sim
\frac 35 \mu$ (free massive fermions), while neutron matter with large scattering lengths carries $E_A/A\sim \frac 34 \mu$
close to the conformal limit.

\section{DIS on extremal RN-AdS}

We now consider DIS scattering on an RN-AdS black hole as the hologram of DIS scattering on a large nucleus at rest
in the double limit of a large number of colors and strong gauge coupling. 
Some useful insights on standard DIS scattering on nuclei can be found in~\cite{REVIEW} to which we refer the interested reader. 
For completeness, we note that DIS scattering on a nucleon using holography was first addressed in~\cite{POL}, and on a thermal black-hole in~\cite{HATTA}.  Although the thermal black-hole is rather different from the extremal RN-AdS black-hole, in the DIS
kinematics they will share much in common as we now detail.

\subsection{Structure functions}

To probe the  RN-AdS black hole in bulk, we use the U(1) R-field ${\bf A}_\mu(x)$ as  the source of the
{\it fermion} bilinear 4-vector current in the boundary of AdS$_5$ ($r=\infty$).  We first use linear response theory to compute the
boundary induced current using an on-shell action. We then relate the  retarded Green function to pertinent structure functions 
in the DIS limit.   The  expectation value of the fermion current is

\be
{\bf J}_\mu(x)=-i\int d^4y\,\left< J_\mu(x)J_\nu(y)\right>_R \,{\bf A}^{\nu}(y)
\label{JJ}
\ee
in the linear response approximation. Here $R$ refers to the retarded correlation function
in the state of finite density.  In Fourier space (\ref{JJ}) simplifies

\be
{\bf J}_\mu(q)=G^R_{\mu\nu} (q)\,{\bf A}^\nu(-q)
\label{RRF}
\ee
with the retarded Green's function

\be
G^R_{\mu\nu}(q)=-i\int\,d^4y\,e^{iq\cdot y}\left<J_\mu(y)J_\nu(0)\right>_R
\label{GREEN}
\ee

The rest frame of the RN-AdS black hole selects the fixed 4-vector $n^\mu=(1,0,0,0)$. 
Current conservation and covariance yields (\ref{GREEN}) in terms of two-invariants
$R_{1,2}$

\be
\label{A.1} 
G^R_{\mu\nu}(x_A,q^2)&&=
\left(\eta_{\mu\nu}-\frac{q_\mu q_\nu}{Q^2} \right)R_1(x_A, q^2)
\nonumber\\&&+\left(n_\mu-\frac{n\cdot q}{Q^2}q_\mu\right)\left(n_\nu-\frac{n\cdot q}{Q^2}q_\nu\right)\,R_2(x_A, q^2)\nonumber\\
\ee
The rest frame of the black hole is the rest frame of the nucleus with fixed energy $E_A=\frac 34 A\mu$. 
Since the binding energy in a nucleus is small, we also have $E_A\simeq Am_N$ and therefore $\mu\simeq \frac 43 m_N$.
In our mapping, $m_N$ and $\mu$ are  interchangeble for estimates.  A hard photon
with virtual momentum $q^\mu= (\omega, 0,0, q)$ scattering off the nucleus in the DIS kinematics satisfies 
$q^2-\omega^2\equiv Q^2\rightarrow\infty$ with $\omega\simeq q$ and  fixed Bjorken-x

\be
\label{XA}
x_A=\frac{q^2}{-2q\cdot (nE_A)}\equiv \frac{Q^2}{2E_A\omega}=\frac{xm_N}{E_A}
\ee
Kinematically, we expect $0\leq x_A\leq 1$ or equivalently $0\leq x\leq A$. 
The DIS structure functions of the RN-AdS black hole will be identified from
the imaginary part of the retarded response function

 \be
 \label{F12}
 2\pi F_1={\rm Im}R_1\qquad 2\pi F_2=\frac{\omega}{E_A}\,{\rm Im}R_2
 \ee

\begin{figure}[!htb]
 \includegraphics[height=10cm]{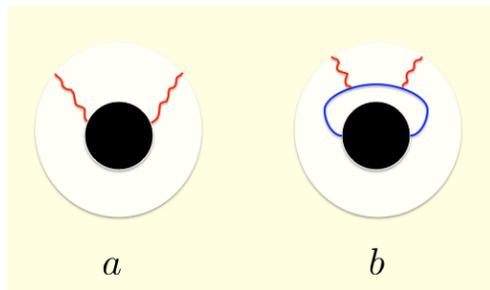}
  \caption{Absorptive virtual-photon  scattering on a nucleus as an extremal RN-AdS black hole: (a) absorptive tree contribution; 
  (b) absorptive one-loop contribution. See text.}
  \label{data}
\end{figure}

\subsection{On-shell action}

To assess (\ref{GREEN}),  we evaluate the metric perturbation induced by the R-current $J_\mu$ in bulk.
The corresponding gravitational wave is vector-like  $A_m(t,\bm{x},u)$ ($m=\mu, u$) and obeys Maxwell
equations in AdS$_5$ with pertinent boundary condition at the boundary ($u=0$) and the horizon ($u=u_h$). The retarted
response function (\ref{GREEN}) is then extracted from the induced action  ${\cal S}[A]$ as a functional of the  boundary fields
$A_\mu(t,\bm{x},0)$ using 

 \beq 
 \label{GRD}
 G^R_{\mu\nu}(q)\,=\,\frac{\partial^2 \mathcal{S}_R}{\partial A_\mu
 \partial A_\nu}\bigg|_{{A_\mu}=A_{{\mu}(u=0)}}\bigg.
 \label{SR} \eeq
The first few leading contributions to (\ref{GRD}) 
are  illustrated in Fig.~\ref{data}a. The first and leading contribution 
is of order  $N_c^2$. This contribution accounts for coherent scattering on the
nucleus as a whole and is dominant at low-x. At intermediate-
and large-x, the subleading correction shown in Fig.~\ref{data}b is more appropriate.
The solid line refers to DIS scattering from a fermion emitted-absorbed by the black-hole.
It is of order $N_c^0$ and more computationally involved~\cite{ZAHED}.
Suffices to say that the emission-absorption of the fermion carries information on the black-hole
as a cold Fermi surface~\cite{MANY}, which can also be addressed using general arguments
along with holography~\cite{ZAHED2}.

 (\ref{RNADS})
yield Maxwell equations in the geometry of the extremal RN-AdS black hole. 
In the gauge $A_u=0$, we dial  the incoming perturbation as a plane wave
with 4-momentum $q^\mu=(\omega,0,0,q)$,

   \be
   \label{pw}
   A_\mu(t,{x},u)\,=\, e^{-i\omega t+iq z}\,A_\mu(u) 
   \ee
   and satisfying  ($i=1,2$)
  
   \beq
    \label{13} 
   \varpi A^\prime_0+kf_{0}A_3^\prime \,=\,0 \label{11} \nonumber\\[0.2cm]
   A_i^{\prime\prime}+\frac{f_{0}^\prime}{f_{0}}A_i^\prime +
   \frac{\varpi^2-k^2f_{0}}{uf_{0}^2}A_i\,=\,0 \label{ai} \nonumber\\
   A_0^{\prime\prime}-\frac{1}{uf_{0}}(k^2A_0+\varpi k A_3)\,=\,0 
   \ee
The prime stands for  a $u$--derivative, and the dimensionless enery and momentum are

 \beq
 \label{dimko}
\varpi\equiv \frac{\omega}{2\pi\gamma\mu}\,, \qquad  k\equiv \frac{q}{2\pi\gamma\mu}\,.
 \eeq
 For the extremal black-hole,
we have re-written the metric (\ref{MADS}) using $u=(r_\pm/r)^2$,

 \beq 
 \label{met1} 
 \rmd s^2=\frac{(\pi \gamma\mu R)^2}{u}(-f_{0}(u)\rmd t^2+\rmd
 \bm{x}^2)+\frac{R^2}{4u^2f_{0}(u)}\rmd u^2\,, \nonumber\\
 \eeq
with  $f_{0}(u)=1-3u^2+2u^3$. The extremal black hole horizon $u=u_{h}=1$ solves $f_{0}(u_{h})=0$.

There are no exact solutions to the wave equations (\ref{ai}). Following~\cite{HATTA} 
we will consider the kinematical limit $k\gg K$  with $K^2=k^2-\varpi^2$, which corresponds
to the low-x regime with $x\ll \mu/Q\sim m_N/Q$. The equation of motion 
can now be solved for  $u\ll u_{h}$,  with the approximate warping 

\be
\label{WARPX}
f_{0}(u)=1-\bigg(\frac{u}{\overline{\gamma}^2}\bigg)^2+\mathcal{O}(u^3)
\ee
and $\overline{\gamma}^{4}=\frac{1}{3}$.  
In this approximation, the warping factor (\ref{WARPX}) for
the RN-AdS black-hole becomes similar to that of an ordinary thermal AdS black hole except for the
differences in scaling with the chemical potential instead of the temperature. At this stage, our analysis
of the longitudinal and transverse waves is similar to the one presented in~\cite{HATTA}  
to which we refer for completeness.

In this regime, (\ref{13}) are solved by sourcing the fields at the AdS boundary, e.g.

\be
\label{LONG}
k(k+A_0+\varpi A_3)(u=0)=k^2{\cal A}_L(0)
\ee
and similarly for the transverse wave ${\cal A}_T$, and by requiring 
absorptive boundary conditions for the ${\cal A}_{T,L}$ waves at the black-hole horizon. 
As a result, the induced boundary action ${\cal S}[A]$ develops  large imaginary parts 
($c$ is Euler's constant)

  \beq
  \label{She} 
  \mathcal{S}_R=&&-\frac{1}{\alpha}\frac{N_c^2\gamma^2\mu^2}{48}
  \Big[k^2\mathcal{A}_L^2(0)
  \left(2\Big(c+\ln\frac{k}{3\overline{\gamma}^2}\Big)-i\pi\right)\nonumber\\
  &&+
  \,\frac{9\pi}
  {\Gamma^2(\frac 13)}
  \left(\frac{k}{3\overline{\gamma}^2}\right)^{\frac 23}
  \left(\frac{1}{\sqrt{3}}-i\right)
  \calbfA_T^2(0) \Big]\,\nn
  \eeq
Modulo the overall constant in (\ref{She}) and the rescaling by $\overline\gamma$, 
the result is  in agreement with the one
derived in~\cite{HATTA} for a non-extremal thermal black hole. We now show how to 
use (\ref{She}) for extracting the  nuclear structure functions at low-x.

\section{Holographic nuclear structure functions}

The holographic structure functions (\ref{F12}) are obtained by inserting 
(\ref{She}) into (\ref{SR}), taking the derivatives and identifying the imaginaty parts.
The result is


  \beq
  \label{FTL} 
  F_T(x_A,Q^2)&=&C_{T}\,\frac{\mu^2}{x_A}\,\left(\frac{x_A^2Q^2}{\mu E_A}
  \right)^{\frac 23}\nn[0.2cm]
  F_L(x_A,Q^2)&=&C_{L}\frac{E_A}{\mu}\,\frac{\mu^2}{x_A}\,\left(\frac{x_A^2Q^2}{\mu E_A}
  \right)
 \eeq
with
\be
&&C_{T}=\frac{N_{c}^2}{2^{17/3}\pi^{2}\Gamma^2(1/3)\alpha^{5/3}}\nonumber\\
&&C_{L}=\frac{N_{c}^2}{1152\pi^4\alpha^2}
\ee
For $x_A\ll \sqrt{\mu E_A}/Q$, we have  $F_L\ll F_T$, which is reminiscent of the Callan-Gross relation
$F_2=F_L+F_T\simeq 2x_AF_1$, noted also for a thermal black hole~\cite{HATTA}. We recall
that at intermediate-x, the structure functions on a spin-$\frac 12$ target obey instead $F_2=2F_1$~\cite{POL}. 
 From (\ref{FTL}) we identify the nucleus saturation line

 \be
 \label{XSQ}
 Q_{AS}(x_A)=\frac {\sqrt{\mu E_A}}{x_A}=\frac {\sqrt 3}2\frac {\mu}{x_A}A^{\frac 12}
 \ee
for  the extremal RN-AdS black hole (\ref{EOS2})  identified as a large nucleus (ignoring binding). 
The saturation momentum  grows with atomic number  $Q_{AS}\sim A^{\frac 12}$ since $x_A\sim A^0$.
In weak coupling, simple QCD arguments  for DIS  suggest $Q_{AS}\approx A^{\frac 13}$. 
 For $A=1$ it is consistent with the thermal saturation line derived in~\cite{HATTA} with the chemical potential
 traded for temperature.




 \subsection{Sum rule}

The range of validity  of (\ref{FTL}) is limited to low-x, parametrically far from the saturation line
for large $A$.  To see this, it is useful to relate the leading twist in the OPE expansion of
 the $JJ$ correlator space-like to the moment of the structure function. For that, we note that in the deep Euclidean regime
 with $q^2\rightarrow\infty$ and $x_A\rightarrow \infty$, the leading twist contribution to $JJ$ is the twist-2 and protected
 energy-momentum operator $T_{\mu\nu}=(\eta_{\mu\nu}+4n_\mu n_\nu)\epsilon/4$. Since $R_{1,2}$ are analytic in the complex 
 $z_A=1/x_A$-plane minus the cuts along $|z_A-1|\leq 0$, we can relate the deep Euclidean region around $z_A\approx 0$ to
 the physical region along the cuts by a Cauchy-transform. This procedure is standard, and the result is the sum rule involving the
 twist-2 operator
 
 \be
 \label{MOMENT}
  \epsilon=&&18E_A^2\int_0^1dx_A\, F_2(x_A, Q^2)\nonumber\\
  \approx &&18E_A^2\,\left(x_AF_2(x_A, Q^2)\right)_{x_A\approx \mu/AQ}
\ee
 The integral is dominated by the low-x region  $x_A\approx \mu/AQ\ll \mu\sqrt{A}/Q$ far from the saturation line
 (\ref{XSQ}).

\subsection{Normalization}

For a comparison with conventional structure functions in DIS on a finite nucleus, we need to address the 
issue of normalization. Indeed, as defined through (\ref{A.1}-\ref{F12}), the holographic structure functions
have dimensions mass-square while the standard ones are dimensionless. The reason is that in  scattering
off the extremal black-hole the state was normalized to 1 instead of  a large  nucleus at rest or 
\be
\label{FACTOR}
(2\pi)^3\, 2E_A\,\delta(\vec 0_p)\equiv \, 2E_A\, V_A \rightarrow (12\pi \alpha)^2\frac {A^2}{N_c^2\mu^2}
\ee
where the rightmost relation follows using the mapping to the  extremal black hole equation of state (\ref{EOS2}).
When inserted in (\ref{FTL}) this factor yields dimensionless structure functions for DIS scattering on a cold nucleus
viewed as an extremal AdS black hole. Inserting (\ref{FACTOR}) into (\ref{FTL}) and using (\ref{XA}) to trade $x_A$ for $x$,
we obtain the properly normalized structure functions at low-x

  \be
  \label{NEW} 
F^A_T(x,Q^2)&=&\tilde C_{T}\,\frac{A}{x}\,
\left(\frac{3x^2Q^2}{4m^2_N}\right)^{\frac 23}\nonumber\\
F^A_L(x,Q^2)&=&\tilde C_{L}\,\frac{3A}{4x}
\left(\frac{3x^2Q^2}{4m^2_N} \right)
 \ee
 with $\tilde C_{T,L}/C_{T,L}=\pi^5{(48\alpha)^2}/{2N_{c}^2}$.

 \subsection{R-ratio}

 In holography, the structure of the nucleon at low-x is dominated by  a virtual photon scattering off 
 a spin-$\frac 12$ dilatino in bulk through a t-exchange of a Pomeron
either  as a surface-exchange~\cite{SURFACE},  or graviton-exchange with the result~\cite{GRAVITON}
 
 \be
 \label{GRA}
 F^N_2(x)=\frac{C_\Delta }{x^{\Delta_{\mathbb P}}}\left(\frac{4m_N^2}{3Q^2}\right)^{\Delta -2}
 \ee
Here $\Delta=mR+2$ refers to  the conformal dimension of the spin$\frac 12$, and the Pomeron intercept 
is $\Delta_{\mathbb P}=2|1-\Delta^2|/\sqrt{\lambda}\ll 1$, with empirically  $\Delta_{\mathbb P}\approx 0.08$.
For $\Delta=\frac 72$, the structure functions obey conformal scaling,  and the corresponding hard form factors  
satisfy the parton-counting rules~\cite{POL}.  Using (\ref{NEW}) and (\ref{GRA}), the nuclear R-ratio follows

 \be
 \label{RATIORA}
 R[x]\equiv \frac {\frac 1A F^A_{2}}{F^N_{2}}=&&\frac{\tilde C_{T}}{C_\Delta}\,
\,\frac{ x^{\Delta_{\mathbb P}+\frac 13}}{x_{S}^{2\Delta-\frac 83}}
\,\bigg(1+ \frac{ 3\tilde C_{L}}{4\tilde C_T}\,\left(\frac x{x_{S}}\right)^{\frac 23}\bigg)\nonumber\\
 \ee
 with $x_{S}\equiv 2m_N/\sqrt{3}Q$.
To compare with the experimentally measured structure functions at low-x in the shadowing region, we need to 
correct  (\ref{RATIORA}) by a surface contribution that is due to the finite size of the nucleus. Recall that the RN-AdS
black hole occupies all of the 3-volume, which is not the case for a dense and large nucleus. This is readily done through

\be
\label{RAC}
R[x]\rightarrow R[x]+\frac {C}{A^{\frac 13}}
\ee
where $C$ is a parameter that cannot be fixed by our arguments.

\begin{figure}[!htb]
 \includegraphics[height=50mm]{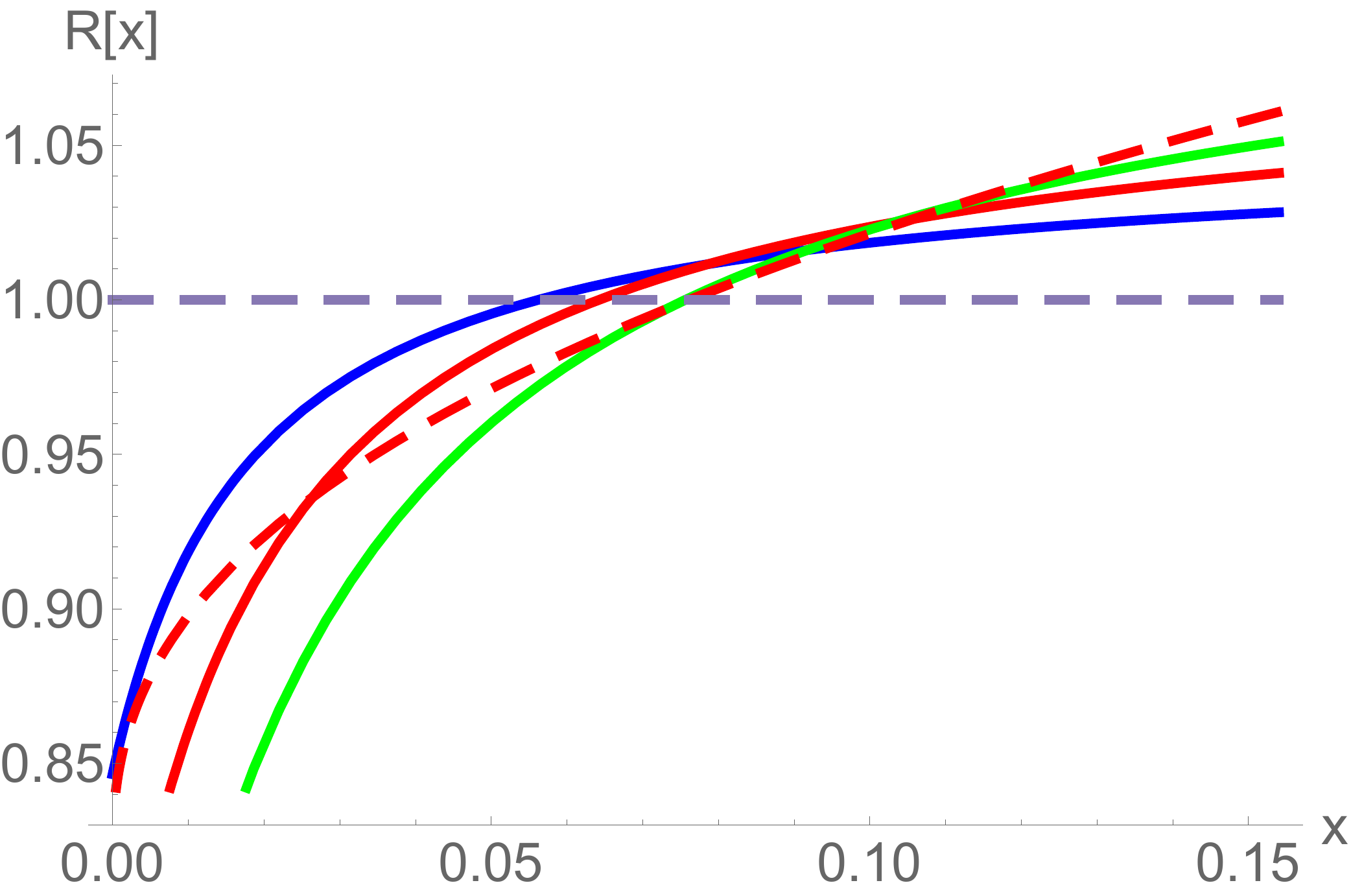}
  \caption{Parametrized DIS data on nuclei (solid curves)  vs holography~(\ref{RATIO}) (dashed curve) in the shadowing region.  See text. }
  \label{shadowing}
\end{figure}

In Fig.~\ref{shadowing} we show the surface corrected 
ratio (\ref{RATIORA}-\ref{RAC})  in the small-x regime ($x\lsim x_S$) as the red-dashed line,
with $A=42$ and the parameters  $C, C_{T,L}/C_{\Delta}$  fixed as

 \be
 \label{RATIO}
0.52\,x^{\frac 13+0.08}+ \frac {2.85}{A^{\frac 13}}
 \ee
 The solid curves are the HPC parametrization of the available nuclear parton distributions from~\cite{PARA}. 
 The upper blue-solid line is for $A=12$, the red-middle line is for $A=42$ and the green-solid line
 is for $A=208$. The holographic estimates support shadowing of the low-x structure functions in
 DIS scattering on an RN-AdS black hole as a model for a dense nucleus.

\section{Conclusions}

We have suggested that DIS scattering on  a large nucleus is analogous to DIS scattering on an extremal RN-AdS black hole 
in holographic QCD. In leading order the absorptive part of the scattering amplitude is dominated by coherent scattering
on the bulk black hole, with structure functions that are dominant at low-x.  We have mapped those results onto a finite nucleus with
fixed atomic number by suitably correcting for an overall normalization. The R-ratio of the structure functions was shown to
exhibit strong shadowing at low-x, an illustration of the strong depletion at low-x through absorption on the black-hole. 
In a way DIS scattering on the RN-AdS black-hole is the ultimate illustration of coherent scattering on a nucleus. 
The effect of Fermi motion at large-x is absent in our leading order analysis. It arises from a subleading DIS scattering 
on the fermions emitted and then absorbed quantum mechanically by the surface of the black-hole. It will be 
addressed next.

\section{Acknowledgements}
This work is supported by the U.S. Department of Energy under Contract No.
DE-FG-88ER40388.




 \vfil

\end{document}